\documentclass[aps,twocolumn,showpacs]{revtex4}
\usepackage[dvips]{graphicx} 
\usepackage{amsfonts,wasysym} 
\usepackage{psfrag} 
\newcommand{\sref}{Sec.~\ref} 
\newcommand{\fref}{Fig.~\ref} 
\newcommand{\tref}{Tab.~\ref} 
\newenvironment{indented}{\begin{center}}{\end{center}}
\newcommand{\br}{\hline}
\newcommand{\mr}{\hline}
\newcommand{\ack}{\paragraph*{Acknowledgments.}}

\usepackage[ps2pdf]{hyperref} 
\hypersetup{colorlinks=false, pdfpagemode=UseOutlines,
 bookmarksopen=false, bookmarksnumbered=true, pdfstartpage= {1},           
 pdftitle    = {Critical Kauffman networks under deterministic asynchronous update},
 pdfsubject  = {Submitted to NJP.}
 pdfauthor   = {florian.greil<at>physik.tu-darmstadt.de},
 pdfkeywords = {Institut fuer Festkoerperphysik, Technische Universitaet Darmstadt, Germany},
 pdfcreator  = {Adobe-Acrobat-Distiller},
 pdfproducer = {LaTeX with hyperref}
}
\newcommand{\Lcm}{{\rm lcm}} 
\newcommand{\LoopCross}{$\Circle \!\! - \!\! \Circle$} 

\begin{document}
\title{Critical Kauffman networks under deterministic asynchronous update}
\author{Florian~Greil, Barbara~Drossel and Joost~Sattler}
\address{Institut f\"ur Festk\"orperphysik,  TU Darmstadt,
Hochschulstra{\ss}e 6, 64289 Darmstadt, Germany }
\date{June 26$^{th}$, 2007} 
\pacs{89.75.Hc, 05.65.+b, 75.10.Nr}
\begin{abstract}
We investigate the influence of a deterministic but non-synchronous
update on Random Boolean Networks, with a focus on critical networks.
Knowing that ``relevant components'' determine the number and length
of attractors, we focus on such relevant components and calculate how
the length and number of attractors on these components are modified
by delays at one or more nodes. The main findings are that attractors
decrease in number when there are more delays, and that periods may
become very long when delays are not integer multiples of the basic
update step. 

\end{abstract} 
\maketitle

\section{Introduction}
Random Boolean networks are widely used as models for complex systems
that consist of interconnected units that influence each other and
have two states (``on'' and ``off''). Kauffman \cite{kauffman:metabolic,
kauffman:homeostasis} used them as a simple model for gene regulation,
but they can also be applied in a social and economic context
\cite{alexander:random, paczuski:self-organized}, for neural networks
and protein networks \cite{kauffman:random}.  Recently it was shown
that the idealized representation of genes as Boolean units is
sufficient to understand the essential dynamics of certain real gene
regulation networks. In these cases, one need not include rate
equations for the concentrations of the molecules involved in the
processes in order to identify the sequence of steps taken by such a
system \cite{bornholdt:less,li:yeast,albert:topology}.

A random Boolean network (RBN) is a directed graph consisting of $N$
nodes and $kN$ links between them. The nodes have values~$\sigma_i \in
\{ 0, 1 \}$ and receive input from $k$ other nodes, which are chosen
at random when the network is constructed. Each node $i$ has an update
function $f_i$, which assigns to each of the $2^k$ states of its $k$
input nodes an output 1 or 0. The update function of each node is
chosen at random among all $2^{2^k}$ possible update function.  All
nodes are usually updated in parallel according to the rule
\begin{eqnarray}
\sigma_i(t+1) &=& f_i(\{\sigma_j(t)\})\, .
\end{eqnarray}
The assignment of connections and functions to each node remains fixed
throughout the whole time evolution, the model is therefore referred
to as ``quenched'' \cite{aldana-gonzalez:boolean}.

The dynamics follows a trajectory $\vec{\sigma}(t) \equiv \{\sigma_1(t),
\ldots, \sigma_N(t)\}$ in configuration space which eventually leads to
a periodically repeating sequence of configurations, called
\emph{cycle}, as the state space is finite and the dynamics is discrete.
Such an cycle is called \emph{attractor} if there is a
set of transient states leading to it; these constitute the
\emph{basin of attraction}.

The dynamics can be classified \cite{aldana-gonzalez:boolean} according
to the way information spreads through the network. 
\begin{enumerate}
\item In the \emph{frozen phase}, all nodes apart from a small number
      (that remains finite in the limit of infinite system size) assume
      a constant value after a transient time.  If in the stationary
      state the value of one node is changed, this perturbation
      propagates during one time step on average to less than one other
      node.
\item In the \emph{chaotic phase} initially similar configurations
      diverge exponentially. Attractors are usually long, and a
      non-vanishing proportion of all nodes keep changing their state
      even after long times. 
\end{enumerate}
\emph{Critical} networks are at the boundary between the frozen and
chaotic phase \cite{aldana-gonzalez:boolean}, and neighboring
configurations diverge only algebraically with time.  Whether a
network is frozen, critical or chaotic depends on the value of $k$ and
on the probabilities assigned to the different types of update
functions.  When all update functions are chosen with the same
probability, networks with $k<2$ are frozen, networks with $k=2$ are
critical, and networks with $k>2$ are chaotic \cite{derrida:random}.

The usual synchronous way of updating is not very realistic
\cite{harvey:time} as natural systems are rarely controlled by an
external clock. It is known that properties of attractors for
synchronous dynamics can differ from those for asynchronous
dynamics. For instance, for cellular automata part of the
self-organization is closely tied to the synchronous updating
\cite{ingerson:structure}.  For RBNs, the dynamics changes
considerably when other updating schemes are chosen
\cite{klemm:stable,klemm:robust}. While in critical Boolean networks
with parallel update the number of attractors increases
superpolynomially with the network size \cite{mihaljev:scaling}, it
becomes a power law for asynchronous stochastic update
\cite{greil:dynamics}.

In this paper we will consider deterministic updating schemes that are
not fully synchronous. This means that some nodes are less frequently
updated than others. Such node-based delays can be motivated
biologically: The expression of genes is not an instantaneous process,
the transcription of DNA and transport of enzymes may take from
milliseconds up to a few seconds.
To each node~$i$ we assign a delay time~$\tau_i$.  The
value~$\sigma_i$ of node $i$ is updated in time intervals $\tau_i$,
and each node must be assigned an initial ``phase''~$\varphi_i <
\tau_i$ (i.e., the time until the first update).  The model is
referred to as a Deterministic Random Boolean Network (DRBN).  The
system is deterministic as the succession of network
states~$\vec{\sigma}(t)$ is entirely defined by the initial
condition~$\vec{\sigma}(0)$ and the initial phases~$\{\varphi_i\}$.
The case of parallel update, the so-called Classical RBN (CRBN), is a
special DRBN with all $\tau_i \equiv 1, \varphi_i \equiv 0$.  The size
of the state space~$\Omega$ changes from $|\Omega|_{\rm CRBN}=2^N$ to
$|\Omega|_{\rm DRBN}=\prod_{i} 2^{\tau_i}$, when all $\tau_i$ are
integers.

The outline of the rest of this paper is the following: First, we
review the concept of relevant components (\sref{RelComp}). This
shows that the most frequent relevant components are simple loops, and
less frequent are collections of loops with additional links within
and between them. In the subsequent sections, we therefore study
simple loops with one delayed node, simple loops with several delayed
nodes, two loops with a cross-link and one delayed node, and a loop
with one additional link and a delayed node
(\sref{LoopsWithOneDelay}-\sref{LoopWithAnInterconnection}).  In
the conclusion, we discuss the consequences of our findings for the
entire network, which is composed of several relevant components.

\section{Relevant components} \label{RelComp}
It has proven useful to classify the nodes of a RBN according to their
behavior on attractors 
\cite{mihaljev:scaling,bilke:stability,socolar:scaling,samuelsson:superpolynomial,drossel:on,kaufman:scaling,bastolla:modular}.
\begin{enumerate}
\item The state of \emph{frozen nodes} becomes constant after some
      time.  Interestingly, the nodes that become frozen are the same
      nodes most of the time, and they constitute the \emph{frozen
      core}. The frozen core is identified by starting from nodes with
      constant functions and by iteratively identifying nodes that
      become frozen due to frozen inputs. Networks without frozen
      functions can also develop a frozen core \cite{paul:properties},
      however, the mechanism is different.  In critical networks, the
      frozen core comprises all but a proportion $\sim N^{-1/3}$ of
      nodes.
\item \emph{Relevant nodes} are non-frozen and have different dynamics
      on different attractors, and they influence at least one relevant
      node \cite{bastolla:relevant}.  They determine the attractors, and
      the number of relevant nodes scales in critical networks as
      $N^{1/3}$ \cite{kaufman:scaling}.
\item Finally, there are the \emph{irrelevant nodes} which are not frozen but
      are slaved by the relevant nodes.  
\end{enumerate}
The non-frozen nodes of
      critical networks essentially form a $k=1$ network. This means
      that typically all but one input of a nonfrozen node are frozen.
In critical networks, relevant nodes are arranged in $\mathcal{O}(\ln
N)$ \emph{components}, most of which are simple
\emph{loops}. Typically, there is only one component that is not a
simple loop but has $\mu$ nodes with two relevant inputs
\cite{kaufman:on}.  
There exist two possible components with $\mu=1$, and then more
complicated components with $\mu>1$.  If $\mu=1$, there are either two
loops with an cross-link or a loop with an additional link, see
\fref{ComponentCartoon}. Two loops with a cross-link occur twice as
often as a loop with an additional link \cite{kaufman:on}.  Since the
relevant components determine the long-term dynamics, we will in the
following study their properties.

\begin{figure}[htb]
\begin{indented}
\item[] 
\includegraphics[width=0.45\textwidth]{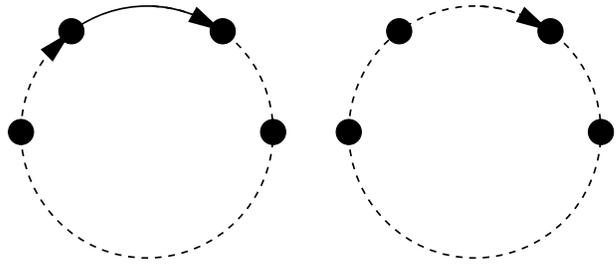}
\caption{The two possible simple loops.  The dashed lines depict a
$\oplus$-coupling, the solid lines a $\ominus$-coupling.  The left one
is an odd, the right one an even loop.}
\label{LoopsSimple}
\end{indented}
\end{figure}

\section{Loops with one delayed node} \label{LoopsWithOneDelay}
The simplest relevant component is a loop consisting of nodes with
$k=1$ incoming edges and Boolean functions which either copy ($\oplus$)
or invert ($\ominus$) the previous node's value. A constant function
within a loop will freeze the whole loop, and can therefore not occur
in relevant loops. A loop with $n$ inversions can be mapped
bijectively onto one with $(n-2)$ inversions by replacing two
$\ominus$ with two $\oplus$ and by inverting the values of all nodes
between these two couplings.  It is therefore sufficient to
distinguish loops with an even or an odd number of inversions, and we
call them ``even'' and ``odd'' loops respectively. When discussing
even loops, we consider loops with only $\oplus$ functions. To odd
loops we assign one $\ominus$ function, the connection to a node with
this function is called ``twisted''.  If no node is delayed, an even
loop with a prime number~$N\in \mathbb{P}$ of nodes returns to its
initial state after $N$ time steps.  If $N$ is not prime, shorter
cycle lengths exist. Irrespective of the updating scheme, there are two
fixed points for an even loop, namely $\vec{\sigma} \in \{ \vec{0},
\vec{1} \}$.  An odd loop with $N\in \mathbb{P}$ nodes returns to its
initial state after $2N$ time steps.  It has no fixed points, For
synchronous update, the shortest attractor of an odd loop has period 2,
with alternating 1's and 0's.

Let us now introduce a delay in the loop. Let node~$1$ be the delayed
node, i{.}e{.}, we choose a value $1<\tau_1 \equiv \tau^\ast \in
\mathbb{N}$, while $\tau_i= 1$ for $i>1$. This means that one gene
needs much longer to be expressed than all the others.  Since node~1
remains at the same value for $\tau^\ast$ time steps, node 2 receives
$\tau^\ast$ times the same input, leading to blocks of size
$\tau^\ast$ travelling around the loop.  When the head of a block
arrives at node $N$, node $N$ will have the value of this block for
$\tau^\ast$ time steps, and during one of these steps node 1 will be
updated. In the following, we will consider even and odd loops
separately.

\paragraph*{Even loops.} 
An even loop with one delayed node has two fixed points, just as the
loop with no delayed node. The other attractors are characterized by
blocks of length~$\tau^\ast$ traveling around the loop. Assume node 1
is updated at time 0. A block with the value of node 1 will start
travelling around the loop, and the head of the block will arrive at
node $N$ at time $N-1$. The next update of node 1 will be at time $T =
\lfloor (N-1+\tau^\ast)/\tau^\ast \rfloor \cdot \tau^\ast$, where the
Gaussian brackets $\lfloor x \rfloor$ denote the largest integer less
or equal $x$. The value of node~1 becomes the same as at time~0, and
the same block travels around the loop again. The same consideration
can be made for all starting times that are multiples of $\tau^\ast$,
leading to the result that the state of the loop is repeated every $T$
time steps.

After the transient time $N-1$, the loop has reached an attractor that
contains $\xi = T/\tau^\ast$ blocks, each of which either has only
values~0 or values~1, and with the blocks that contains node~1 and $N$
being shorter than the other ones, if $N$ is not a multiple of
$\tau^\ast$.  Every attractor corresponds to a pattern of blocks
travelling around the loop. If the number of blocks is a prime number,
$\xi \in \mathbb{P}$, the length~$A_\oplus$ of the attractors is
identical to $T$,
\begin{eqnarray}
A_\oplus &=&T =\left\lfloor \frac{N-1+\tau^\ast}{\tau^\ast}\right\rfloor
\cdot \tau^\ast=\xi \cdot \tau^\ast \qquad \forall \, \xi \in \mathbb{P}\,.
\end{eqnarray}
 For $\tau^\ast \ge N$ there is only $\xi=1$ block containing
all nodes, and the fixed points are the only attractors.

For $\xi \notin \mathbb{P}$, the pattern of blocks can have a period
that is a divisor of $\xi$, in which case the attractor length is
shorter.

The number of different attractors, $\nu_\oplus$, can be calculated from
the number of different patterns, $2^\xi$. Including the 2~fixed points
this leads to
\begin{eqnarray}
\nu_\oplus &=& \frac{2^\xi-2}{\xi} +2 \qquad \forall \, \xi \in \mathbb{P}
\end{eqnarray}
If the number of blocks is not prime, $\xi \notin \mathbb{P}$, the
number of attractors increases as the length of some attractors is
shorter. 

\paragraph*{Odd loops.} 
Without loss of generality we assign the twisted edge of the odd loop to
be in front of the delayed node. As in the synchronous case, there are
no fixed point attractors for odd loops. Let again node~1 be delayed and
updated at time~0.  At time $T$, node~1 will have the opposite state as the
original one.  After time $2T$, node 1 returns to its original state,
which implies that the loop returns to its original state after $2T$
time steps, if it is on an attractor. If all nodes are identical
initially, a single domain wall travels around the loop, and after $T$
time steps all nodes are again identical, but with the opposite state.
The shortest attractor has a period $2\tau^\ast$, and it has alternating
blocks.  If the number of blocks $\xi = T/\tau^\ast$ is a prime number,
all other attractors have the period $2T$, and the number of different
attractors is 
\begin{eqnarray}
\nu_\ominus &=& \frac{2^\xi-2}{2\xi} +1\, .
\end{eqnarray}
If $\xi \notin \mathbb{P}$, the number of attractors increases as the
length of some attractors is shorter.  If $\tau^\ast \ge N$, there is
only one attractor with period $2\tau^\ast$. 

\paragraph*{Non-integer delays.} 

Let us now consider the case that $\tau^*$ is not an integer, but a
rational or a real number. Real numbers can be approximated by a
series of rational numbers, and we therefore consider the case $\tau^*
= r/s$ with two incommensurate integers $r$ and $s$, with
$r>s$. During $r$ time steps, $s$ blocks emerge from node 1, part of
them of length $\left\lfloor \frac r s \right\rfloor $, part of them
of length $\left\lfloor \frac r s \right\rfloor +1 $.  When the first
block reaches node 1, the same sequence of blocks will emerge again
only if node 1 is in the same phase at its next update as it was at
its first update. Otherwise, the pattern of blocks will be changed at
each circulation around the loop, until it starts repeating again
after $s$ (or $2s$ for an odd loop) circulations (or a divisor of
it). It follows that for irrational values of $\tau^*$ the dynamics
never become exactly periodic but are quasiperiodic. Of course, for
values $\tau^*>N$, the only attractor is a fixed point (for an even
loop) or a state with only one domain wall (for an odd loop).

\section{Loops with multiple delayed nodes}
Next, we consider loops with multiple delayed nodes and integer delay
times $\tau_i \in \mathbb{N}_{>0}$.  Loops with rational values
$\tau_i=r_i/s_i$ can be mapped on those with integer values of
$\tau_i$ by measuring time in units of the inverse of the least common
multiple of all $s_i$. 
In the following we will first look at two special cases before we focus
on the general case where updates may occur in any order. 

\paragraph*{Sequential update.} 
We choose $\tau_i \equiv N$ and update the nodes in the order in which
they occur on the loop, i.e., $\varphi_i = i-1$ (connection-wise (cw)
update) or $\varphi_i = N-i$ (counter-connectionwise (cc) update).

For cw-update, all nodes of an even loop have the same value, after
every node has been updated once. Thus, we have two fixed point
attractors consisting of the two homogeneous configurations,
$\vec{\sigma} \in \{\vec{0},\vec{1}\}$.  For an odd loop, the
attractor has a single domain wall that travels around the loop, and
the period of the attractor is $2N$. 

For cc-update, node $j$ is updated before node $j-1$. Therefore, $N$
update steps give the same result as 1 update step in the case of
parallel update, and the results of \sref{LoopsWithOneDelay} can be
taken over.

\paragraph*{Same delay, different phases.} 
We now choose again $\tau_i \equiv N$, but we update the nodes in any
order, i.e., the values of $\varphi_i$ are some permutation of the
numbers 0 to $N-1$.  There are two classes of nodes, according to
whether a node is updated before or after its predecessor. 
Nodes that are updated after their predecessor have after $N$ time
steps the same state as their predecessor. Such nodes and their
predecessor are therefore part of the same \emph{effective node}. The
number $N^\ast$ of effective nodes is identical to the number of nodes
that are updated after their predecessor. 
Let us give an example,  for instance $N=6$ and the updating order
\begin{eqnarray*}
\{\varphi_i\} &=& \{5,0,3,1,2,4\}
\end{eqnarray*}
 If we specify only whether a node
is updated before (b) or after (a) its predecessor, this can be
written as
\begin{eqnarray*}
 && \{a,b,a,b,a,a\}
\end{eqnarray*}
leading to $N^\ast =2$. 
Once we have identified the effective nodes, we can map $N$ time steps
 on such a loop with sequential update on one time step on a loop of
 size $N^\ast$ with parallel update. All results concerning attractor
 numbers and lengths obtained for loops with parallel update can then
 be transferred to loops with sequential update.

\paragraph*{Different delay times.} 
We now consider the general case where the delays~$\tau_i$ and the
phases~$\varphi_i$ can take any integer value.
In order to determine whether the initial state of a given node
influences the attractor, we proceed in the following way: We fix the
state of this node, let us say, to 1, and we evaluate to which nodes
this 1 propagates with time. In order to make sure that later on all 1s
on the loop will be due to this initial 1, we set all other nodes to 0
and choose an even loop.  When the chosen node is updated before its
successor, the 1 is lost, and the initial state of this node does not
affect the attractor. If the node is updated after its successor, the 1
has moved to the successor and is not yet lost. Next, we check whether
the successor is updated before the 1 that is now there can propagate
further. During the course of time, the 1 may spread to become a block
of larger size, which continues to change its size with time. Now we
consider the loop at times which are a multiple of $\tau^\ast =
\Lcm(\tau_i)$. At these times, the phases of all nodes are the same as
at the beginning.  We wait until either the original node has again
state 1 or until all 1s are gone. In the first case, the 1 will survive
forever, and the initial state of the chosen node will consequently
affect the attractors. In the second case, the chosen node does not
affect the attractors. 

If we repeat this procedure for every node, we will know the initial
state of how many  nodes $N^\ast$ will affect the attractors. We only
consider these ``relevant'' nodes from now on, and we consider them only
at times that are multiples of $\tau^\ast$. Let $m$ be the number of
relevant nodes through which each block moves during $\tau^\ast$ time
steps. If $m$ and $N^\ast$ have no common divisor, we order the $N^\ast$
nodes in the sequence in which they are visited if the system is only
considered at times that are multiples of $\tau^\ast$. Then we have
mapped the task of finding the attractor number and length on a loop of
size $N$ with any rational delay times of the task of finding the
attractor number and length in a loop of size $N^\ast$. If $m$ and
$N^\ast$ have a common divisor $l$, we can map our system on $l$ loops
of length $N^\ast/l$ with parallel update and thus find the number and
lengths of attractors. 

For irrational delay times, the mapping on integer delays cannot be
performed. Nevertheless, one can determine whether a 1 at a given
node will eventually become a block that is so large that it will
never vanish. If all delays have irrational ratios, there will
eventually come a moment where all nodes are updated
connection-wise, and from then on there is only one block left.

\section{Two loops with a cross-link}
\begin{figure}[htb]
\begin{indented}
\includegraphics[width=0.45\textwidth]{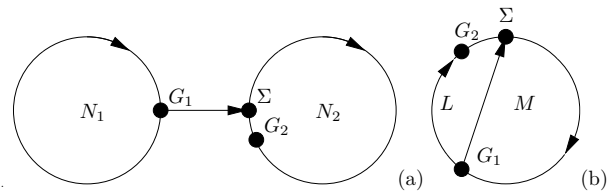}
\caption{\label{ComponentCartoon} The complex components considered in
this paper: (a) The \LoopCross-component consists of a loop with $N_1$
nodes connected to a loop with $N_2$ nodes. (b) The $\oslash$-component
is a loop of $N=L+M+2$ nodes having an additional link.}
\end{indented}
\end{figure}

Now we consider a complex component consisting of a loop with $N_1$
nodes connected to a loop with $N_2$ nodes, see
\fref{ComponentCartoon} (a). The node~$\Sigma$ is the one with two
inputs, and its input nodes are labelled $G_1$ and $G_2$.  The first
loop is either odd or even, the second loop can without loss of
generality be chosen such that it has only $\oplus$-couplings, except
at $\Sigma$. We insert no nodes between $G_1$ and $\Sigma$, as a system
with $m$ nodes on the cross-link can be mapped on a system with a
direct link by connecting node number~$m$ (counted clockwise
from $G_1$) directly with node~$\Sigma$.

We consider only the nontrivial cases where the coupling function of
node~$\Sigma$ is a function that responds to both of its inputs.
If we take into account that certain of these functions can be mapped
onto each other by inverting the states of all nodes or by inverting
the states on the first loop, we end up with three functions that are
truly different. They are shown in \tref{TruthTable}.
\begin{table}
\begin{indented}
\item[] \begin{tabular}{cclccc} 
\br 
\multicolumn{2}{c}{inputs}&&reversible & 
\multicolumn{2}{c}{canalizing functions}\\ \cline{5-6}
$~G_1~$ & $~G_2~$ &&$f_{\rm rev}$& $f_{\rm ci}$ & ~$f_{\rm ch}$ \\ 
\mr 
0&0&&1&1&0\\ 
0&1&&0&0&1\\
1&0&&0&1&1\\ 
1&1&&1&1&1\\
\br \end{tabular} 
\caption{\label{TruthTable} Boolean functions used in the
\LoopCross-component for node~$\Sigma$.  The second letter in the label
of the canalyzing functions stands for ``homogeneous'' or
``inhomogeneous''.} 
\end{indented}
\end{table}

In the following, the results by Kaufman and Drossel \cite{kaufman:on}
for components under synchronous update will be generalized to
components with one delayed node.

\paragraph*{Delayed node on first loop.} 
If the first loop has a delay~$\tau^\ast$, the value of $G_1$ can change
only at times that are multiples of $\tau^\ast$, and the pattern of
change is repeated after the attractor period of the first loop. If the
first loop is on a fixed point, the second loop can be considered as an
independent loop with a function at $\Sigma$, which depends on the fixed
point value of the first loop. We therefore consider here the case that
the first loop is not on a fixed point but provides a periodic input of
period $p_1$ to $\Sigma$, with blocks of size $\tau^\ast$ of identical
bits. The second loop then behaves like an independent loop where the
Boolean function at $\Sigma$ changes after $\tau^\ast$ steps, and where
the pattern of changes is repeated periodically with period $p_1$.

If $f_\Sigma = f_{\rm rev}$, the second loop switches between truth
and negation, for the (in)homogeneous canalizing function $f_{\rm ch}
(f_{\rm ci})$ the second loop changes between truth (negation) and
constant value~$\sigma_\Sigma = 1$. The attractor period is at most
$2p_1N_2$.
More detailed results for possible attractors of such a system under
synchronous updating can be found in \cite{kaufman:on}, the only
difference being that $p_1$ is now related in a different way with
$N_1$. An interesting finding is that for the homogeneous canalizing
function, the second loop becomes frozen on the value 1 if $p_1$ and $N_2$
have no common divisor.  Furthermore, for the inhomogeneous canalyzing
function, the first loop enslaves the second loop and imposes its period
on it, if $p_1$ and $N_2$ have no common divisor.

\paragraph*{Delayed node on second loop.} 
We now proceed to the case where a single delayed node is on the second
loop.  The first loop behaves in the same way as for simple
(synchronously updated) CRBN-loops.  We focus on the sequence of values
of the delayed node. We assign the
delay to the node~$\Sigma$: A system with the delayed node $m$ nodes
after $\Sigma$ can be transformed into a system with the delay at $\Sigma$
by rotating the first loop $m$ nodes counterclockwise.

Node~$\Sigma$ responds to the input from $G_1$ only every $\tau^\ast$
time steps.  Let us denote with $p_1$ the period of the sequence of
values of $G_1$ every $\tau^\ast$ time steps, which is the period of
the input sequence to $\Sigma$ generated by the first loop at those
times where $\Sigma$ is updated.  Let $\xi = \lfloor
(N_2-1+\tau^\ast)/\tau^\ast \rfloor$ denote the number of blocks of
the second loop.

All results for the attractors on two loops with a crosslink and no
delay can now taken over by replacing $N_2$ with $\xi$, by replacing
nodes with blocks, and by taking $\tau^\ast$ as
the time unit.  In particular, for a reversible function $f_\Sigma$,
the largest period is $p_1 \xi \tau^\ast$. A homogeneous canalyzing
function $f_\Sigma$ freezes the second loop on the value 1 if $p_1$
and $\xi$ have no common divisor.  Furthermore, for an inhomogeneous
canalyzing function, the first loop enslaves the second loop and
imposes its period (times $\tau^\ast$) on it if $p_1$ and $\xi$ have
no common divisor.

\paragraph*{General case.} 
Now we consider the case of multiple (integer) delays on both loops.
Let the first loop have a period $p_1$ and the second loop (if even
and decoupled from the first loop) a period $p_2$. The general system
of two interconnected loops without any delay has been studied in
\cite{kaufman:on}. There, it was shown that the attractor length lies
between $p_1$ and $2 p_1 N_2 / g$, where $g = \Lcm(p_1,N_2)$.  We can
conclude that now the attractor length lies between $p_1$ and $ 2 p_1
p_2 / g$, where $g$ is the greatest common divisor of $p_1$ and $p_2$.

For a homogeneous canalyzing function $f_\Sigma =
f_{\rm ch}$, the second loop is frozen if $p_1$ and $p_2$ are
incommensurable. The longest attractors occur for reversible functions,
$f_\Sigma= f_{\rm rev}$.

\section{Loop with one additional link} \label{LoopWithAnInterconnection}
\begin{figure*}[htb]
\begin{indented}
\includegraphics[width=0.95\textwidth]{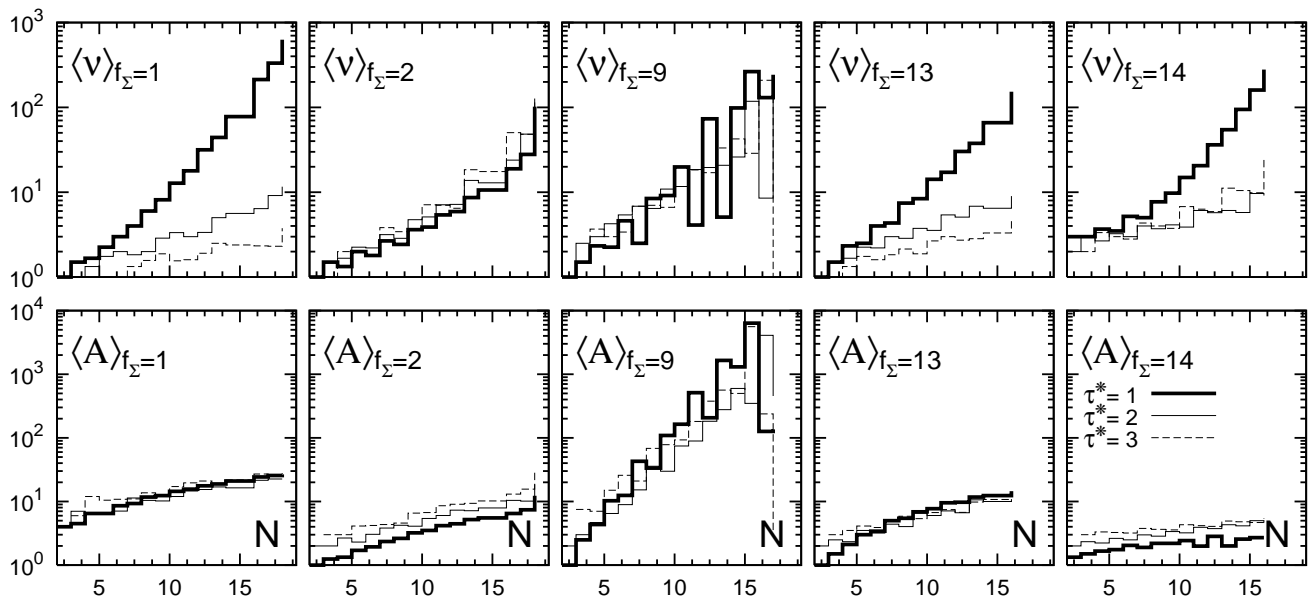}
\caption{Results of the exhaustive state space search for the
$\oslash$-component with small system size, $N<20$.  The upper panel
shows the mean attractor number~$\langle \nu\rangle$ in dependence of
the system size~$N$ while the lower shows the mean attractor
length~$\langle A \rangle$ for fixed delay $\tau^\ast$ and for all
functions $f_\Sigma \in\{1,2,9,13,14\}$. In both cases the average was
taken over all possible realizations which corresponds to an average
over all $L$ for a single delayed node.  To keep the diagrams concise
only some delays have been plotted.} \label{DrbnNxxLxxTxxFxx}
\end{indented}
\end{figure*}

The other complex component with one node with two inputs is a loop
with $N=L+M+2$ nodes and one additional link. We call again the node
with two inputs $\Sigma$, and its inputs $G_1$ and $G_2$, see
\fref{ComponentCartoon} (b). The link from $G_2$ to $\Sigma$ can be
treated as a direct link: A system with $n<L$ nodes in the additional
link can be mapped onto a system with a direct link by connecting
node~$(M+1+n)$ to $\Sigma$ (if we neglect delays).  We consider five
update functions at $\Sigma$, compare \tref{TruthTableTwo} (we now
use the common decimal representation as identifiers for the
functions). The other canalizing or reversible functions yield the
same result, one only has to invert the output values of the truth
table. Without loss of generality, all other functions in the loop are
copy functions.

Let there be one delayed node in the component.  We distinguish two
cases according to the position of the node with update
period~$\tau^\ast>1$:
\begin{enumerate}
\item The delayed node lies on the first $M+1$ nodes (including
$G_1$). Without loss of generality the delay can be shifted to
node~$\Sigma$.
\item The delayed node is in the chain of nodes between $G_1$ and
$\Sigma$ and can then be shifted to~$G_2$.
\end{enumerate}

In the first case, the component can be reduced to a network of
effective nodes by looking at the network only every $\tau^\ast$ time
steps.  Each effective node corresponds to a block of $\tau^\ast$ nodes
which are at the same state for $t~{\rm mod}~\tau^\ast = 0$.  The
results for the synchronous case (as studied in \cite{kaufman:on},
Sec.~4) hold for the effective variables~$\tilde{N},\tilde{M}$, $\lceil
x \rceil$ denotes the smallest integer greater or equal to $x$:
\begin{eqnarray}
\tilde{N} &=& \left\lceil \frac{N}{\tau^\ast}   \right\rceil 
\qquad \qquad 
\tilde{M} = \left\lceil \frac{M+2}{\tau^\ast} \right\rceil -2
\end{eqnarray}
In the following we will study the second case, where $G_2$ is the
delayed node. 

\begin{table}[htb]
\begin{indented}
\item[] \begin{tabular}{@{}cccccccc}
\br $~G_1~$ & $~G_2~$ &&
$f_{\rm 1}$& $f_{\rm 2}$&$f_{\rm 13}$&$f_{\rm 14}$&$f_{\rm 9}$\\
\mr
0 &0 && 1 & 0 & 1 & 0 & 1 \\
0 &1 && 0 & 1 & 0 & 1 & 0 \\
1 &0 && 0 & 0 & 1 & 1 & 0 \\
1 &1 && 0 & 0 & 1 & 1 & 1 \\ 
\br
\end{tabular}
\caption{\label{TruthTableTwo} Boolean functions used in the
$\oslash$-component for node~$\Sigma$.  The names for the functions are
the decimal representation of the corresponding column of outputs, for
instance $1\cdot 2^0+0\cdot2^1+1\cdot2^2+1\cdot2^3=13$. Function $f_9$
is reversible, the other functions are canalyzing.}
\end{indented}
\end{table}

\paragraph*{Canalyzing function at $\Sigma$.} 
If there is a canalyzing function at $\Sigma$, there exist at least
two types of attractors. The first type of attractors is obtained by
requiring that $G_2$ does never have its canalyzing value at the
moment when $\Sigma$ is updated. (The canalyzing value is 0 for
$f_{13}$ and $f_2$ and 1 for $f_{1}$ and $f_{14}$.) In this case the
loop consisting of the $M+2$ nodes from $\Sigma$ to $G_1$ is an even
loop for $f_{13}$ and $f_{14}$ and an odd loop for $f_1$ and
$f_{2}$. Just before node $\Sigma$ is updated, the state of node $G_2$
and the state of all nodes that will in $n\tau^\ast$ time steps
determine the state of $G_2$ must have a value such that $\Sigma$ never
has its canalyzing value ($n$ is any positive integer). For functions
$f_{13}$ and $f_{14}$, this condition fixes the entire component at
the same value if $M+2$ and $\tau^\ast$ have no common divisor. If
their greatest common divisor $g$ is larger than 1, only the value of
every $g^{\rm th}$~node on the loop of length $M+2$ is fixed by this
condition. The number of attractors of the first type is therefore
that of an even loop with $(M+2)(g-1)$ nodes and with no delays.  For
functions $f_1$ and $f_{2}$, the condition that $G_2$ does never have
its canalyzing value can only be satisfied if $\tau^\ast/g$ is even.
The number of attractors of the first type is then that of an odd loop
with $(M+2)(g-1)$ nodes and with no delays.  Compared to a component
with no delays, this new type of attractors increases the mean
attractor length if the canalyzing function is $f_{14}$. Without
delays, only very short attractors (apart from two fixed points) can
occur \cite{kaufman:on} for $f_{14}$, and the increase in attractor
length for values $\tau^\ast>1$ is clearly visible in
\fref{DrbnNxxLxxTxxFxx}.

On all other attractors, $G_2$ has at least sometimes its canalyzing
value. The second type of attractors referred to above is obtained if
$M+2$ is a multiple of $\tau^\ast$. If the loop of $M+2$ sites
consists of blocks of size $\tau^\ast$, the dynamics can be mapped on
that of an effective component with $\tilde{N} = \left\lceil
\frac{N}{\tau^\ast}\right\rceil$ nodes and with $\tilde{M} = \frac{M+2}{\tau^\ast}
-2$ with no delay but time steps of length $\tau^\ast$, and the
results of \cite{kaufman:on} can be taken over. In addition to
attractors consisting only of homogeneous blocks, further attractors
can be constructed by realizing that only one bit in each block is the
one that triggers node $G_2$. The value of bits that do not trigger
node $G_2$ does only matter when the block reaches $G_1$: If at this
moment $\Sigma$ is not canalyzed by $G_2$, an inhomogeneous block will
not be homogenized, but copied or inverted to node $\Sigma$. An
inhomogeneous block can therefore survive forever if the blocks that
are at $G_2$ at the moment where the inhomogeneous block is at $G_1$ do
not have their canalyzing value. However, this implies that these
noncanalyzing blocks are copied to $\Sigma$ from $G_1$, which is only
possible for $f_{13}$.  Indeed, from \cite{kaufman:on} we know that a
period $\tilde M + 2$ of attractors on the effective component is only
possible for this function, unless $\tilde N$ has special values. 

Finally, let us look for nontrivial attractors that can occur even if
$M+2$ and $\tau^\ast$ have no common divisor. Let us choose the
function $f_2$. An isolated block of size $\tau^\ast$ of 1s in the
initial state will survive forever since there will be a 0 at $G_1$
while this block is at $G_2$. In fact, there exist a multitude of such
attractors where there is a 0 at the right position at distance $L$
behind a block of 1s. This explains why the number of attractors found
numerically for $\tau^\ast > 1$ is larger for $f_2$ than for the other
canalyzing functions (see \fref{DrbnNxxLxxTxxFxx}). The length of
these attractors can be larger than $N+\tau^\ast-1$, as can also be
seen in \fref{DrbnNxxLxxTxxFxx}. 

\paragraph*{Function~$f_9$.} 
$f_{9}$ is a reversible function, i{.}e{.}, if one of the inputs changes
its value the output changes, too. If there are no delays, the dynamics
is reversible, and therefore all states are on cycles. There is 1 fixed
point $\vec{\sigma} =\vec{1}$. A striking feature of the synchronous
case is that cycles of the order $2^N$ exist \cite{kaufman:on}. 

The exhaustive numerical attractor search suggests that the maximal
attractor length can be approximated by $\tau^\ast$ times the maximal
attractor length in the non-delayed case for odd $\tau^\ast$.

\section{Conclusion}
In this paper, we studied the influence of a deterministic but
non-synchronous update on critical Kauffman network by introducing
node-based delays. The dynamics of critical networks can be derived
from the dynamics on its relevant components, most of which are simple
loops, and some of which have a few nodes with two inputs. For this
reason, we have studied in this paper the three simplest types of
relevant components. Not surprisingly, delays typically increase the
attractor lengths and reduce the attractor numbers. New types of
attractors emerge in the presence of delays. The basins of attraction are
naturally larger and thus the path to the attractor becomes more
robust. If all delays are randomly chosen real numbers, loops are most
likely to be frozen or on a single attractor. Similarly, more complex
components with real delays should have far less attractors than for
parallel update.

It will be interesting to see how these results are affected when
nonrandom networks, such as real gene regulation networks, are
considered. Clearly, they are not updated in parallel.  Some networks,
such as in budding yeast \cite{li:yeast} appear to be very robust with
respect to the introduction of delays. This means that their choice of
connections and functions is such that the update sequence does not
matter much. It remains to be seen if this is a general feature of all
those networks than can be described by using a Boolean
idealization.

\ack
This work was supported by the Deutsche Forschungsgemeinschaft (DFG) under
Contract No. Dr200/4-1.

\bibliographystyle{utcaps.bst}
\bibliography{fgDrbnBib.bib}
\end{document}